\documentclass[conference]{IEEEtran}
\usepackage{url}
\usepackage{graphicx}
\usepackage{amsmath}
\usepackage{amsfonts}
\usepackage{gensymb}
\usepackage{soul}
\usepackage{color}
\usepackage{bm}
\DeclareMathOperator*{\argmax}{arg\,max}
\newcommand{\xvec}{\operatorname{\bm{{x}}}}

\newcommand{\rbij}{\operatorname{r_{\textit{ij}}^{(\textit{b})}}}
\newcommand{\rpij}{\operatorname{r_{\textit{ij}}^{(\textit{b}^\prime)}}}
\newcommand{\bprime}{\operatorname{\textit{b}^\prime}}

\newcommand{\qEHVI}{$q$EHVI}

\title{Optimizing Coverage and Capacity in Cellular Networks using Machine Learning}
\begin{document}

\author{\IEEEauthorblockN{Ryan M. Dreifuerst\IEEEauthorrefmark{1},
Samuel Daulton\IEEEauthorrefmark{2}, Yuchen Qian\IEEEauthorrefmark{2},
Paul Varkey\IEEEauthorrefmark{2}, Maximilian Balandat\IEEEauthorrefmark{2}, Sanjay Kasturia\IEEEauthorrefmark{2},\\ Anoop Tomar\IEEEauthorrefmark{2}, Ali Yazdan\IEEEauthorrefmark{2}, Vish Ponnampalam\IEEEauthorrefmark{2}, Robert W. Heath\IEEEauthorrefmark{3}}
\\
\IEEEauthorrefmark{2}Facebook, Menlo Park CA\\
\IEEEauthorrefmark{1}Wireless Networking and Communications Group, The University of Texas at Austin, Austin TX\\
\IEEEauthorrefmark{3}Department of Electrical and Computer Engineering, North Carolina State University, Raleigh NC}
\maketitle

\date{\today}

\setlength{\abovedisplayskip}{0pt}
\setlength{\belowdisplayskip}{5pt}
\setlength{\abovedisplayshortskip}{0pt}
\setlength{\belowdisplayshortskip}{0pt}

% Comment these out to remove IEEE copyright notice
\IEEEoverridecommandlockouts
\IEEEpubid{\begin{minipage}{\textwidth}\ \\[12pt]
  \copyright 2020 IEEE.  Personal use of this material is permitted. Permission from IEEE must be obtained for all other uses, in any current or future media, including reprinting/republishing this material for advertising or promotional purposes, creating new collective works, for resale or redistribution to servers or lists, or reuse of any copyrighted component of this work in other works.
\end{minipage}}

\maketitle

\begin{abstract}
Wireless cellular networks have many parameters that are normally tuned upon deployment and re-tuned as the network changes. Many operational parameters affect reference signal received power (RSRP), reference signal received quality (RSRQ), signal-to-interference-plus-noise-ratio (SINR), and, ultimately, throughput. In this paper, we develop and compare two approaches for maximizing coverage and minimizing interference by jointly optimizing the transmit power and downtilt (elevation tilt) settings across sectors. To evaluate different parameter configurations offline, we construct a realistic simulation model that captures geographic correlations. Using this model, we evaluate two optimization methods: deep deterministic policy gradient (DDPG), a reinforcement learning (RL) algorithm, and multi-objective Bayesian optimization (BO). Our simulations show that both approaches significantly outperform random search and converge to comparable Pareto frontiers, but that BO converges with two orders of magnitude fewer evaluations than DDPG. Our results suggest that data-driven techniques can effectively self-optimize coverage and capacity in cellular networks.
\end{abstract}

\begin{IEEEkeywords}
coverage and capacity optimization, Bayesian optimization, machine learning, reinforcement learning
\end{IEEEkeywords}

\section{Introduction}
Simultaneous optimization of coverage and capacity (CCO) is a central problem in the deployment and configuration of wireless cellular networks. Data-driven methods are ideally suited to this problem, given the complexity of network configuration and the availability of performance-related data. However, prior work on intelligent network configuration in industry has only considered rule-based configuration settings, e.g. self-organizing network (SON) paradigm for 4G networks \cite{AliuSON}, and required costly manual oversight due to the many radio environments and deployment configurations \cite{BuenestadoTuning, TemesvaryTuning, cco-1}. Machine learning (ML) and modern Bayesian optimization techniques have the potential to identify solutions that improve performance and reduce the operationally expensive manual oversight.

\IEEEpubidadjcol
As data collection capabilities of cellular networks increase, there are more opportunities to tackle CCO with ML \cite{cco-2}. In addition, without improved optimization, the demand on cellular networks will require dramatic densification \cite{Jurdi-holes}. Densification will increase cell overlap and result in massive, complex networks to manage--an impossible task for rule-based systems. Initially, optimization was only considered for static cell boundaries by first dividing a region into serving cell blocks and then optimizing poorly performing cells using ML techniques \cite{cco-2}. This does not allow cell regions to change, however, which is very inaccurate when antenna downtilt is modified. Reinforcement learning is a useful tool for CCO thanks to its natural formulation as a state-action-reward dynamic (\cite{RazaviRL}, \cite{BaleviRL}). Early work was reported in \cite{RazaviRL}, but this required individual cells to be optimized independently. Multi-cell optimization was performed with RL in \cite{DandanovSO}, but the choice of optimization parameters for signal power and interference power was not investigated. More recently, \cite{BaleviRL} proposed a multi-agent mean field RL algorithm for macrocell optimization. While scalable to large networks, this method did not consider realistic radio channel characteristics.

\IEEEpubidadjcol
In this paper, we propose optimizing capacity and coverage in a multi-cell network by jointly tuning the downtilt and the transmit power in each sector. We treat the coverage and capacity objectives as black-box functions, with no analytical formula and no gradient observations. The goal is to maximize coverage and capacity by optimizing the downtilt and transmit power in each sector over a bounded set $\mathcal X \subset \mathbb R^d$. In this multi-objective optimization problem, there is no single best solution; rather, the goal is to identify the set of Pareto optimal solutions, where any increase in coverage means degrading capacity (and vice versa). We evaluate two black-box optimization techniques for solving the multi-objective optimization problem: Bayesian optimization (BO) \cite{BalandtBotorch}, a sample-efficient method that uses a probabilistic surrogate model to balance exploration and exploitation, and the deep deterministic policy gradient algorithm (DDPG) \cite{Silver2014DeterministicPG}, which has been shown to have strong empirical performance over high-dimensional, continuous action spaces. Since there is no single best solution, we compare the methods in terms of the quality of the solution sets identified by each method and the sample efficiency of each method. Since running field experiments on a real cell network is time-consuming and could degrade network performance, we develop a spatially consistent RF coverage map simulator for evaluating the optimization methods. Due to the computational complexity of realistic propagation models, we combine offline computation of radio propagation using a MATLAB-based open-source simulator from the Fraunhofer Heinrich Hertz Institute \cite{Quadriga} with Python-based post-processing to compute the user-defined coverage and capacity objectives. Our code and simulations are available on Github \cite{github-repo}.

\section{Problem Formulation}
\label{problem_formulation}
\label{pf}
Consider a pre-defined area of interest containing multiple base stations with multiple sectors, under the management of a single controller.  We identify areas of \emph{under coverage} (coverage holes) as those locations that do not have enough received signal strength. We identify areas of \emph{over-coverage} as those where the interfering cells are generating too much interference. To define these concepts more precisely, we use the notion of RSRP. In LTE networks, RSRP is a common performance indicator reported by user equipment (UE) to represent the signal level and coverage quality, which is used for determining the serving cell. Using the RSRP, under covered areas are those locations where the maximum RSRP from any sector falls below a pre-specified threshold while over-covered areas are the locations where the difference between the maximum RSRP and the sum of received powers from other sectors does not exceed another pre-specified threshold. The use of pre-defined thresholds is based on receiver sensitivity, selectivity, network density, and interference management for frequency reuse, which is standard in LTE networks. Common values for these thresholds are 6dB for over-coverage and -110 dBm for under-coverage.

A \emph{configuration} $\xvec$ specifies a joint setting of transmit power and downtilt for each sector. In this work, we consider 10 discretized values for downtilt selection and allow the transmit power to be a continuous value defined over a restricted range. Note that the joint setting space, for $N$ controllable antennas, is exponential in $N$, which precludes naive brute-force search.

Let $\gamma_w$ and $\gamma_o$ be the weak and over coverage thresholds, respectively. We consider a 2D-grid world representation of the area of interest, and index the row and column, for each grid point, using $i$ and $j$, respectively. In the following, we let $b$ denote the sector antenna from which the highest signal power is received at any location, also known as the \emph{attached} or \emph{serving} sector, and $\rbij$ denotes RSRP from that sector. $\bprime$ represents any other sector.

If $\rbij < \gamma_w$ (or, $\rbij - \sum_{b'\neq b}\rpij< \gamma_w$), location [$i, j$] is said to be \emph{under-covered} (respectively, \emph{over-covered}). Minimizing both under-coverage and over-coverage presents a multi-objective optimization problem. Typically in multi-objective optimization, there is no single best solution; rather, the goal is to identify the set of Pareto optimal solutions. A solution is Pareto optimal if any improvement of one metric deteriorates another. Provided with the Pareto set of optimal trade-offs, network operators can select a Pareto optimal configuration according to their preferences. The choice of optimizing over coverage thresholds ensures serving areas are evaluated fairly and network-wide performance metrics can be defined over regions rather than UE specific metrics.

It is illustrative to understand the significance of these two objectives, by inspecting a linear combination of them. Let a linear combination of these objectives depend on setting $x$ and parameter $\lambda$. Then, an optimal setting $\xvec^*$ that maximizes a linear combination may be written as :

\begin{align*}
\xvec^* &= \argmax_{\xvec} u(\xvec) \nonumber \\
    &= \argmax_{\xvec}\lambda \sum_{i,j} \rbij + (1 - \lambda) \sum_{i,j} \bigg( \rbij - \sum_{\bprime\neq b}\rpij\bigg) \nonumber \\
    &= \argmax_{\xvec} \sum_{i,j}\bigg[\rbij - (1 - \lambda) \cdot \sum_{\bprime\neq b}\rpij\bigg] \nonumber\\ %: \text{regrouping}\\
    &\overset{a}{=} \argmax_{\xvec} \sum_{i,j}\bigg[10 \log S_{ij}^{(b)} - (1 - \lambda) 10 \log I_{ij}^{(b)}\bigg] \\ %: \text{change of variables}\\
    &= \argmax_{\xvec} \sum_{i,j}\bigg[10 \log \bigg(\frac{S_{ij}^{(b)}}{(I_{ij}^{(b)})^{(1-\lambda)}}\bigg)\bigg] \nonumber
\end{align*}
We change variables from $r_{ij}^{(b}$ to $10\log S_{ij}^{(b)}$ and from $\sum_{\bprime\neq b}\rpij$ to $10\log I_{ij}^{(b)}$ in ($a$), to signify signal and interference power (in dB).

Thus we see that the optimization seamlessly trades off between optimizing signal to interference power ($\lambda = 0$) and RSRP ($\lambda = 1$), which roughly correspond to optimizing for capacity and coverage, respectively.

In the rest of this work, we select the following two sums of sigmoid functions, centered around the respective thresholds, as the dual objectives to optimize. The under-coverage \eqref{eqn: undercov} and over-coverage \eqref{eqn: overcov} functions are defined as

\begin{align}
    &\sum_{i,j}\sigma\big(\gamma_w - \rbij\big) : \text{under coverage} \label{eqn: undercov}\\
    &\sum_{i,j}\sigma\bigg(\sum_{\bprime\neq b}\rpij - \rbij + \gamma_o\bigg) : \text{over coverage.} \label{eqn: overcov}
\end{align}
where $\sigma$ is the sigmoid function. The use of sigmoids is important because gradient-based optimization with hard thresholding results in sparse gradients and poor learning. Specifically, learning only occurs if a region is either under- or over-covered, but not if a region could be improved. Instead, a soft thresholding function that saturates, is differentiable, and monotonic will lead to dense gradients. The sigmoid function in particular is a commonly used method for soft thresholding. Note that the signs are flipped to signify that both under-coverage and over-coverage are to be \emph{minimized}.

\section{Simulation data generation}
\label{simulation_setup}

\begin{figure}[htb]
	\centering
	\includegraphics[width=1.0\linewidth]{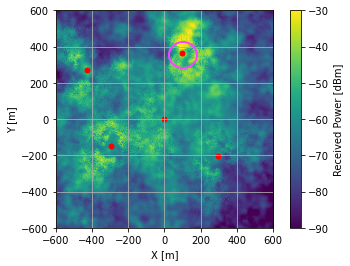}
	\caption{Example of an RSRP map summed over all the sectors. Base stations are marked by a red circle, and all antennas are configured to $5\degree$ downtilt and $46$dBm transmit power. The upper right base station, circled, has a strong effect on local RSRP due to being placed only 20m above ground as opposed to 25-30m for the other sites.}
	\label{fig:rsrp_map}
\end{figure}

The RF Reference Signal Receive Power maps are simulated using a MATLAB tool suite called QuaDRiGa \cite{Quadriga}. We consider a $1.2\times1.2\  \text{km}^2$ network site composed of five base stations with three sector antennas each with controllable downtilt, $d_i \in \{0, 1,...10\}$, and transmit power, $p_i \in [30, 50]$ dBm, making the configuration vector $\xvec = [a_1, a_2... a_{15}]^T,\ a_i = [d_i,\ p_i]^T$. \iffalse{The exact location and azimuth rotation of each sector antenna can be found in Table \ref{tb: sectors}.}\fi Locations of the base stations and azimuth orientations of the sectors are selected in advance and antenna parameters are based on commonly available hardware. Simulating each of the exponentially many joint settings is infeasible. Instead, we sample the RF channel parameters once for each grid point from each sector antenna. Then we iteratively store the coverage map generated by $a_i$ for each downtilt, totaling $150$ coverage maps for one RF environment. During model training, for any given input joint configuration, ${\xvec}$, the associated downtilt maps are scaled by adjusting the RSRP at each point by the associated transmit power. An example of the sum of RSRP maps over each sector antenna is shown in Fig. \ref{fig:rsrp_map}.

\begin{figure}[htb]
    \centering
    \includegraphics[width=1.0\linewidth]{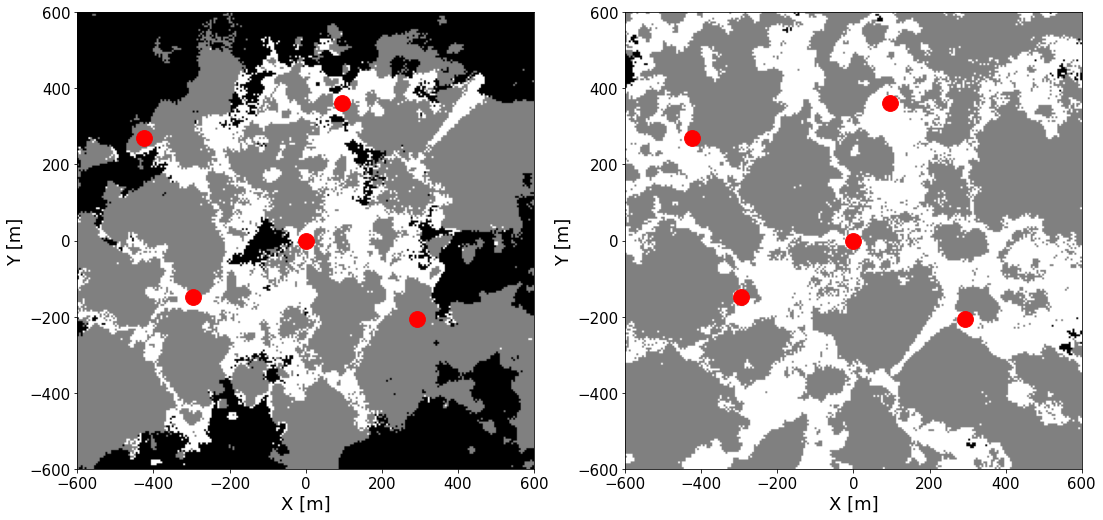}
    \caption{\label{fig:max_min_powers} RF Coverage maps showing under-coverage and over-coverage measures for minimum and maximum powers, respectively. Under-coverage is shown in black while over-coverage is shown in white.}
\end{figure}

To ground intuition about the dual objectives of interest (under-coverage and over-coverage), in Figure \ref{fig:max_min_powers} we illustrate the RF Coverage KPIs for minimum power and maximum power, on the left and right sides, respectively. The downtilt was arbitrarily set. The gray color represents good coverage, and black and white represent, respectively, areas of under-coverage (prevalent in the low-power regime) and over-coverage (prevalent in the high-power regime).

\section{Bayesian Optimization}
\label{bo}
Bayesian optimization (BO) is a sample-efficient method for black-box optimization that that uses a probabilistic surrogate model to balance exploration and exploitation. The surrogate model is typically a Gaussian Process (GP), which provides a posterior distribution $P(\bm f | \mathcal D)$ over the true function values given observed data $\mathcal D = \{\bm x_n, \bm y_n\}_{n=1}^N$ and is known for its well-calibrated uncertainty estimates \cite{Rasmussen2004}. BO provides a distinct advantage over deep learning approaches because of the uncertainty estimates, and natural evolution from Bayes Theory. In BO, the next candidate is selected as the configuration that maximizes an acquisition function $\alpha(\bm x)$ that provides the utility of evaluating $\bm x$ on the true function. While the true function is expensive-to-evaluate, the acquisition function is cheap-to-compute and optimize using the surrogate model to select the next candidate $\bm x_{N+1} = \argmax_{\bm x} \alpha(\bm x)$.

Evaluating a configuration $\bm x$ in a real deployment is time-consuming because it requires running a field trial using $\bm x$ and measuring the observed weak-coverage and over-coverage. Since only a single configuration can be evaluated at a time, efficiently converging to the set of optimal configurations is critical.

We model each objective with an independent GP with Matern-$5/2$ ARD kernel and fit the GP hyperparameters using maximum a posteriori estimation. In order to optimize both objectives simultaneously, we use the $q$-Expected Hypervolume Improvement (\qEHVI) acquisition function \cite{daulton2020differentiable}. We initialized the optimization with space-filling design using 512 points from a scrambled Sobol sequence and then ran 500 iterations of BO.

\section{Reinforcement Learning : DDPG}
\label{ddpg}
The DDPG algorithm integrates deep neural networks with the deterministic policy gradient algorithm. Based on the actor-critic architecture, it takes advantage of policy-based and value-based learning methods to learn `end-to-end' policies in high-dimensional and continuous action spaces \cite{Silver2014DeterministicPG}. The actor network learns the policy mapping states to specific actions deterministically and the critic network learns $Q$-values of action-state pairs.

\begin{figure}[htb]
    \centering
    \includegraphics[width=0.85\linewidth]{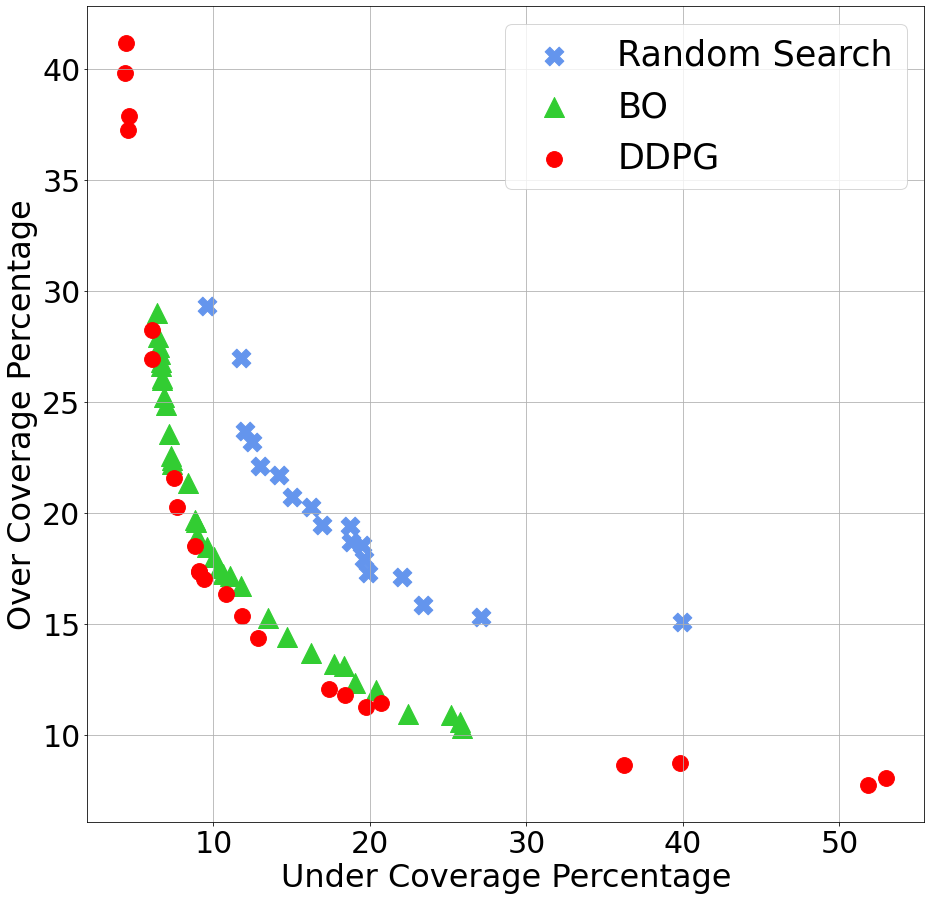}
    \caption{\label{fig:ddpg_vs_random} Comparison of the Pareto frontier for random search, BO, and DDPG. DDPG achieves the best frontier, with an average improvement of $1.0\%$ over BO. As expected, the two conflicting metrics create a convex frontier in $\mathbb{R}^2$.}
\end{figure}

In the current DDPG model, the states are defined as the operational status of different sectors. Typically all the sectors will work fine, and states remained unchanged during the time. The actions contain the $N$ length configuration vector, $\xvec$. The reward function is defined to be a convex combination of two coverage metrics: $$\lambda \sum_{i,j}\sigma\big(\gamma_w - \rbij\big) + (1 - \lambda) \sum_{i,j}\sigma\bigg(\sum_{\bprime\neq b}\rpij - \rbij + \gamma_o\bigg).$$ We formulate the reward as a convex combination because the two metrics, over- and under-coverage are intuitively and visually shown to be convex in the BO results from Fig. \ref{fig:ddpg_vs_random}. As a result, a convex combination will explore the Pareto frontier.
The algorithm will concentrate on areas of over-coverage or areas of under-coverage according to different $\lambda$ values (see discussion in Section \ref{problem_formulation}).

The simulation goes through $\lambda$ settings from 0 to 1 with 0.1 strides, to determine the best configurations in different situations. The replay memory capacity is set to 5,000. At training time, explorations are performed by adding Gaussian noise with decaying variance to the actions. Initially, the noise variance starts at 1, and the decay coefficient is 0.9996 per iteration. For one specific setting $\lambda$, the algorithm will run for 30,000 iterations. The choice of parameters did not greatly affect performance, but we found these to perform best in our investigation.

\begin{figure}[htb]
    \centering
    \includegraphics[width=1.0\linewidth]{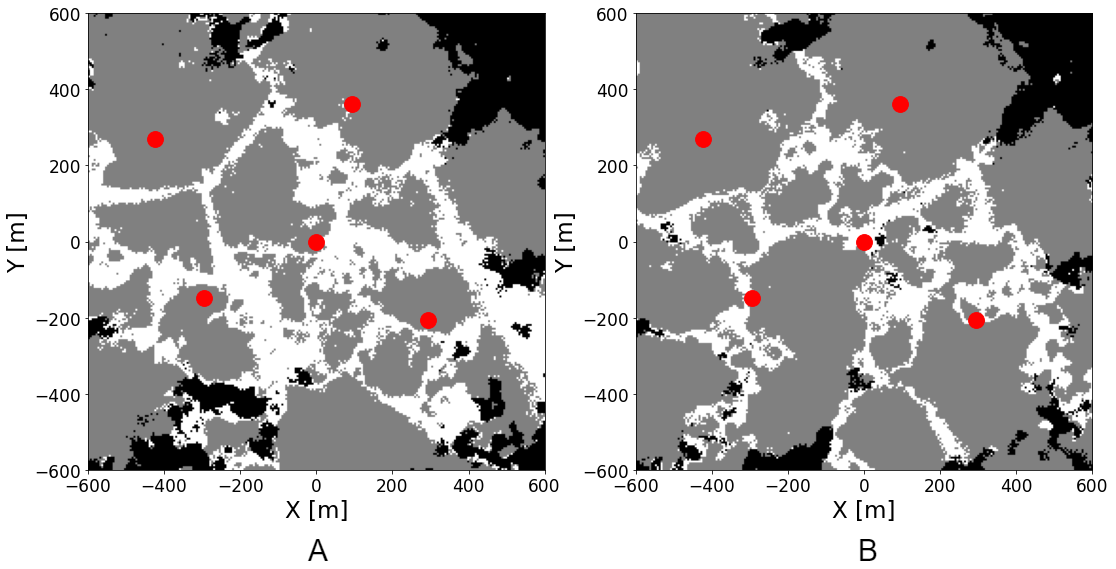}
    \caption{\label{fig:ddpg-rf}Random vs. DDPG: When $ \lambda = 0.6$, (A) the RF Coverage map with the best settings from random search resulting in 12\% under-coverage and 25\% over-coverage area (B) the RF Coverage map with optimized settings from DDPG search resulting in 9\% under-coverage and 17\% over-coverage area.}
\end{figure}

Figure \ref{fig:ddpg-rf} contrasts the improved nature of RF coverage obtained by DDPG (right-hand side) over the best setting randomly found (left-hand side) for $\lambda = 0.6$ (as before, gray represents areas of good coverage, while black and white represent under-coverage and over-coverage, respectively). Note that the results from BO are visually similar and so are not shown here.

\section{Discussion and future work}
In Figure \ref{fig:ddpg_vs_random}, we compare DDPG and BO with a baseline policy that randomly selects configurations with uniform  probability and retains the best configuration. Both BO and DDPG significantly outperform random search and identify comparable Pareto frontiers. In finding the frontier, however, DDPG required an explicit convex combination, whereas BO is able to sample along the frontier through multi-objective optimization. We also see that the RL approach is able to identify a slightly better Pareto frontier. However, DDPG required evaluating over two orders of magnitude more configurations than BO: the DDPG Pareto frontier was identified with 300,000 evaluations and BO only used 1,012 evaluations. Given that BO can identify a comparable Pareto front and improve sample efficiency by over two orders of magnitude relative to DDPG, \emph{BO shows significant promise for optimizing network coverage and capacity in real-world field experiments where sample efficiency is crucial.}

In future work, we consider two directions: network scaling and risk-averse optimization. Scaling up to larger geographic areas with larger search spaces will be critical for real world deployment, where optimization from a single central source may be infeasible for hundreds or thousands of cells. Another practical concern is ensuring that a parameter configuration does not lead to unacceptable quality of network service (e.g. from large coverage gaps). Neither the DDPG nor BO methods considered in this work ensure that only "safe" parameter configurations will be tested, but there there is a body of literature on risk-averse BO \cite{cakmak2020bayesian} and RL \cite{Geibel2012} that could be used to mitigate the risk of severe degradation in network performance.

Automated coverage and capacity optimization (CCO) in cellular networks has a long history beginning with rule-based or heuristic SONs. With the extensive data collection capabilities now supported by 3GPP, LTE networks are now moving from rule-based SONs to ML-based SONs \cite{cco-2}. Our work considers two approaches: deep deterministic policy gradient and Bayesian optimization. We have created a simulation environment to generate realistic RF environments and have shown the results of these algorithms for optimizing cell configurations to balance the conflicting goals of capacity and coverage in a network. Our results demonstrate the utility of principled multi-objective optimization for sample-efficient network optimization.

\bibliographystyle{IEEEbib}
\bibliography{refs}

\end{document}